\documentclass[reprint,aps,prl,superscriptaddress]{revtex4-1}

\pdfoutput=1

\usepackage{graphicx}
\usepackage[dvipsnames]{xcolor}
\usepackage{calc} 

\usepackage{amsmath}
\usepackage{amssymb}
\usepackage{bm}

\usepackage{braket}
\usepackage{units}

\usepackage{eurosym}

\usepackage{url}
\usepackage[colorlinks=true,
        linkcolor=black,
        citecolor=black,
        filecolor=black,
        urlcolor=black,
        bookmarks=false,
        bookmarksopen=false,
        bookmarksopenlevel=3,
        plainpages=false,
        pdfpagelabels=true,hyperfootnotes,
        pdfauthor={J. A. Mlynek, A. A. Abdumalikov Jr, J. M. Fink, L. Steffen, M. Baur, C. Lang, A. F. van Loo, A. Wallraff},
        pdftitle={Time Resolved Collective Entanglement Dynamics in Cavity Quantum Electrodynamics}
        ]{hyperref}

\begin{document}


\title{Demonstrating W-type Entanglement of Dicke-States\\
in Resonant Cavity Quantum Electrodynamics}


\author{J.~A.~Mlynek}
\altaffiliation{These authors contributed equally to this work.}
\author{A.~A.~Abdumalikov Jr}
\altaffiliation{These authors contributed equally to this work.}
\author{J.~M.~Fink}
\author{L.~Steffen}
\author{M.~Baur}
\author{C.~Lang}
\author{A.~F.~van Loo}
\author{A.~Wallraff}
\affiliation{Department of Physics, ETH Z\"urich, CH-8093, Z\"urich, Switzerland.}
\date{\today}

\begin{abstract}
Nonlinearity and entanglement are two important properties by which physical systems can be identified as non-classical. We study the dynamics of the resonant interaction of up to N=3 two-level systems and a single mode of the electromagnetic field sharing a single excitation dynamically.
We observe coherent vacuum Rabi oscillations and their nonlinear $\sqrt{N}$-speed up
by tracking the populations of all qubits and the resonator in time. We use quantum state tomography to show explicitly that the dynamics generates maximally entangled states of the W class in a time limited only by the collective interaction rate. We use an entanglement witness and the threetangle to characterize the state whose fidelity $F= 78\%$ is limited in our experiments  by crosstalk arising during the simultaneous qubit manipulations which is absent in a sequential approach with $F = 91\%$.
\end{abstract}

\maketitle

Effects related to the interaction of light and matter can be investigated in many different physical systems. The description of this interaction can be simplified to the interplay of identical two-level systems and a single mode of an electromagnetic field. This theoretical abstraction is known as the Dicke- or Tavis-Cummings model \cite{Dicke1954, Tavis1968}. In recent years it became feasible to investigate the complex interaction of light with multiple two-level systems as experiments approached a higher level of control over collections of individual quantum systems. The coupling strength (transition amplitude) of N two-level systems and a single mode increases as $\sqrt{N}$.
Clear evidence of this nonlinear behavior was observed in spectroscopic measurement with few atoms \cite{BERNARDOT1992,Childs1996,Thompson1998,Munstermann2000}, large ensembles of atoms using cold gases \cite{Brennecke2007,Colombe2007,Tuchman2006}, ion Coulomb crystals \cite{Herskind2009} and by using superconducting qubits coupled to a transmission line resonator \cite{Fink2009, Altomare2010a}.
The investigation of these interactions has also gained additional momentum in the context of hybrid quantum systems, in which ensembles of microscopic systems, such as NV-centers are coupled to a single mode of a cavity \cite{Kubo2010,Wu2010a}.

More insight into the dynamics of collective systems can be gained by time resolved measurements of energy exchange between its individual components. When multiple two-level systems are resonantly coupled to a single mode, this process is called collective vacuum Rabi oscillations. The collective coupling strength $g$ defines the frequency of these oscillations. Here, we restrict our investigation to the initial state in which the cavity is populated with exactly one photon. In this case the oscillation involves a single photon that is continuously absorbed and reemitted by all N two-level systems. In fact each two-level system absorbs the photon with equal probability 1/N. Our lack of knowledge about which two-level system actually absorbs the photon leads to entangled states which are known as W-states \cite{Duerr2000}. The nonlinear enhancement of the coupling strength speeds up the generation of W-states when N is increased. In the context of quantum information processing such enhanced collective interaction rates may prove useful to generate multi-qubit entangled states on time scales $\propto 1/\sqrt{N}$. Alternatively, the generation of Dicke- and W-states has been explored in NMR \cite{teklemariam2002}, with photons \cite{eibl2004}, ions \cite{roos2004} and superconducting circuits \cite{Neeley2010a} using interactions not mediated by the resonant interaction with a cavity.

Here, using quantum state tomography, we demonstrate explicitly that the strong collective coupling mediated by the resonant interaction of $N$ superconducting qubits with a single photon stored in a transmission line resonator generates W-type entangled states as predicted by the Dicke and Tavis-Cummings Hamiltonians \cite{Dicke1954,Tavis1968}. This work builds on the prior spectroscopic \cite{Fink2009} and time resolved \cite{Altomare2010a} observations of the $\sqrt{N}$ enhancement of the collective resonant interaction with the cavity. 3-qubit entangled states of the W- or GHZ-class have also been generated in superconducting circuits using sequential \cite{DiCarlo2010,Baur2012,Fedorov2012} or collective schemes \cite{Neeley2010a} based on physical interactions different from the one presented here.

In our time-domain studies of the collective oscillation we use
full control over three transmon-type superconducting qubits \cite{Koch2007}
coupled to a single electric field mode of a microwave resonator. The
coupling of multiple two-level systems to a single mode is described by
the Tavis-Cummings Hamiltonian \cite{Tavis1968}
\begin{equation}\label{tch}
\hat{\mathcal{H}}_{\textrm{TC}}= \hbar\omega_\textrm{r}
\hat{a}^\dagger \hat{a} + \sum_{j}{\left(
\frac{\hbar}{2}\omega_j\hat{\sigma}^{z}_j
+\hbar g_j(\hat{a}^\dagger \hat{\sigma}^-_j+
\hat{\sigma}^+_j\hat{a} )\right)} \, .
\end{equation}
$\omega_\textrm{r}$ is the frequency of the field mode, $\omega_j$ is the transition frequency between ground state $|g\rangle$ and excited state $|e\rangle$ of the two-level system $j$, $\hat{a}$ and $\hat{a}^\dagger$ are the creation and annihilation operators of the mode and $\hat\sigma^\pm_j$ are the corresponding
operators acting on the qubits.
\begin{figure}[!b]
  \centering
  \includegraphics[width=1.0\columnwidth]{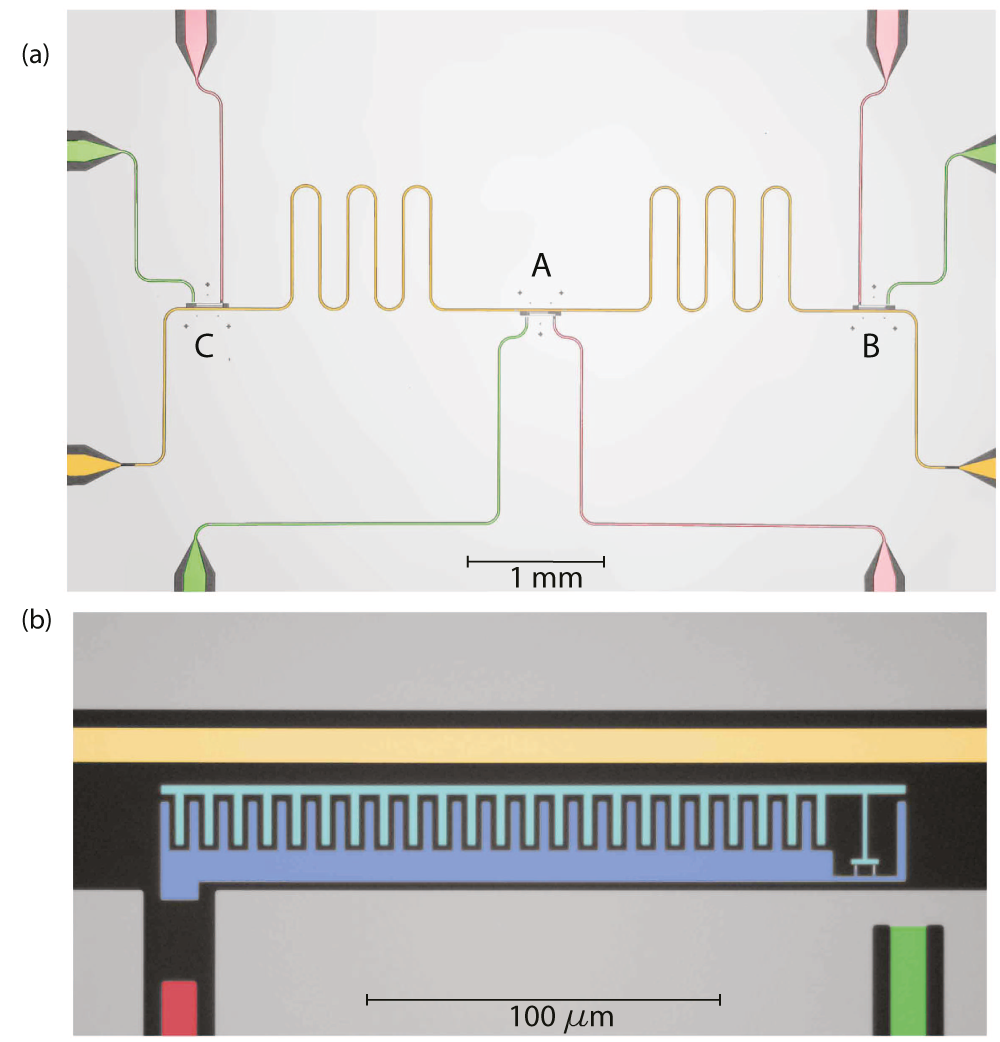}
  \caption{(Color online) (a) Optical microscope false color image of the sample with three qubits (A,B,C) capacitively coupled to a coplanar waveguide resonator (yellow). Each qubit is equipped with individual local charge (red) and magnetic flux-bias lines (green).  (b) Enlarged view of the transmon qubit~C. The island (blue) and the reservoir (lighter blue) are connected via a SQUID loop.}
  \label{fig:sample}
\end{figure}
In our circuit QED implementation of the Hamiltonian $\hat{\mathcal{H}}_{\textrm{TC}}$ we use the first harmonic mode of a coplanar transmission line resonator at $\omega_r/2\pi\simeq$ 7.023~GHz with quality factor $Q\simeq14800$. A false color optical microscope image of the sample is shown in Fig.\,\ref{fig:sample}(a/b). The resonator is used for the resonant exchange of a single excitation and for joint dispersive readout of the three qubits by measuring its transmission \cite{Filipp2009b}. All qubits are located at antinodes of the first harmonic mode of the resonator. The frequency of each qubit is approximately given by $\hbar\omega_j(\Phi)\approx\sqrt{8 E_C E_J (\phi)}-E_C$ where the Josephson energy depends periodically on the applied flux according to $E_J(\phi)=E_{Jmax}|\cos(\pi\phi/\phi_0)|$. The maximum Josephson energies $E_J(0)/\hbar$ for the three qubits are (26.8, 28.1, 25.7) GHz and their charging energies $E_C/\hbar$ are (459, 359, 358) MHz. The anharmonicity of the transmon energy levels depends on $E_C$ and is chosen such that the validity of the two-level approximation is ensured while keeping the charge dispersion low. The maximum transition frequency of the qubits $\omega_j(0)/2\pi=(9.58, 8.65, 8.23)$~GHz is designed such that all three qubits can be tuned into resonance with the resonator at $\omega_r$. In the steady state the qubits are flux biased at $\omega_j/2\pi\simeq(6.11, 4.97, 7.82)$~GHz using the quasistatic magnetic field generated by three superconducting miniature coils positioned underneath the chip, such that $|\Delta_j|\gg g_j$ with detuning $\Delta_j=\omega_j-\omega_r$. We measure qubit dephasing times $T_2$ = (100, 140, 440) ns limited by flux noise far off the optimal bias point and qubit relaxation times $T_1$ = (2.1, 1.8, 1.0) ${\mu}$s limited by unknown reasons other than the Purcell-effect. Tuning of the qubit transition frequencies on the nanosecond timescale is achieved by injecting current pulses into on-chip flux control lines [Fig.\,\ref{fig:sample}(b)]. Both for the coils as well as for the flux gate lines we determined the full coupling matrix to compensate for cross-coupling at low frequencies. The coupling strengths $g_j$ of the qubits (A,B,C) extracted from spectroscopic measurements are $(g_A,g_B,g_C)/\pi$ = $(-105.4, 110.8, 111.6)$~MHz. The negative coupling constant of qubit~A originates from the $\pi$-phase difference of the first harmonic mode between the center of the resonator and its coupling ports.

By applying phase controlled truncated Gaussian DRAG pulses \cite{Gambetta2011a} through the on-chip charge bias lines [Fig. \ref{fig:sample}(b)] each qubit can be prepared in an arbitrary superposition state $\psi=\alpha\ket{g}+\beta\ket{e}$. For the joint read-out all qubits are dispersively coupled to the resonator and its measured frequency-shift depends on the state of each individual qubit as $Tr(\hat M\rho)$, where $\hat M$ is the measurement operator and $\rho$ is the density matrix to be determined \cite{Filipp2009b}. To extract the 64 unknown elements of the density matrix we have applied 64 linearly independent measurement operators on the state to be characterized. These independent operators are constructed by applying a set of single qubit rotations, namely $\{Id,~\sigma_x,~\sigma_y,~\sigma_z\}$ to the operator $\hat M$. To extract the population only, the number of linearly independent operators reduces to 8.

\begin{figure*}
  \centering
  \includegraphics[width=1.0\textwidth]{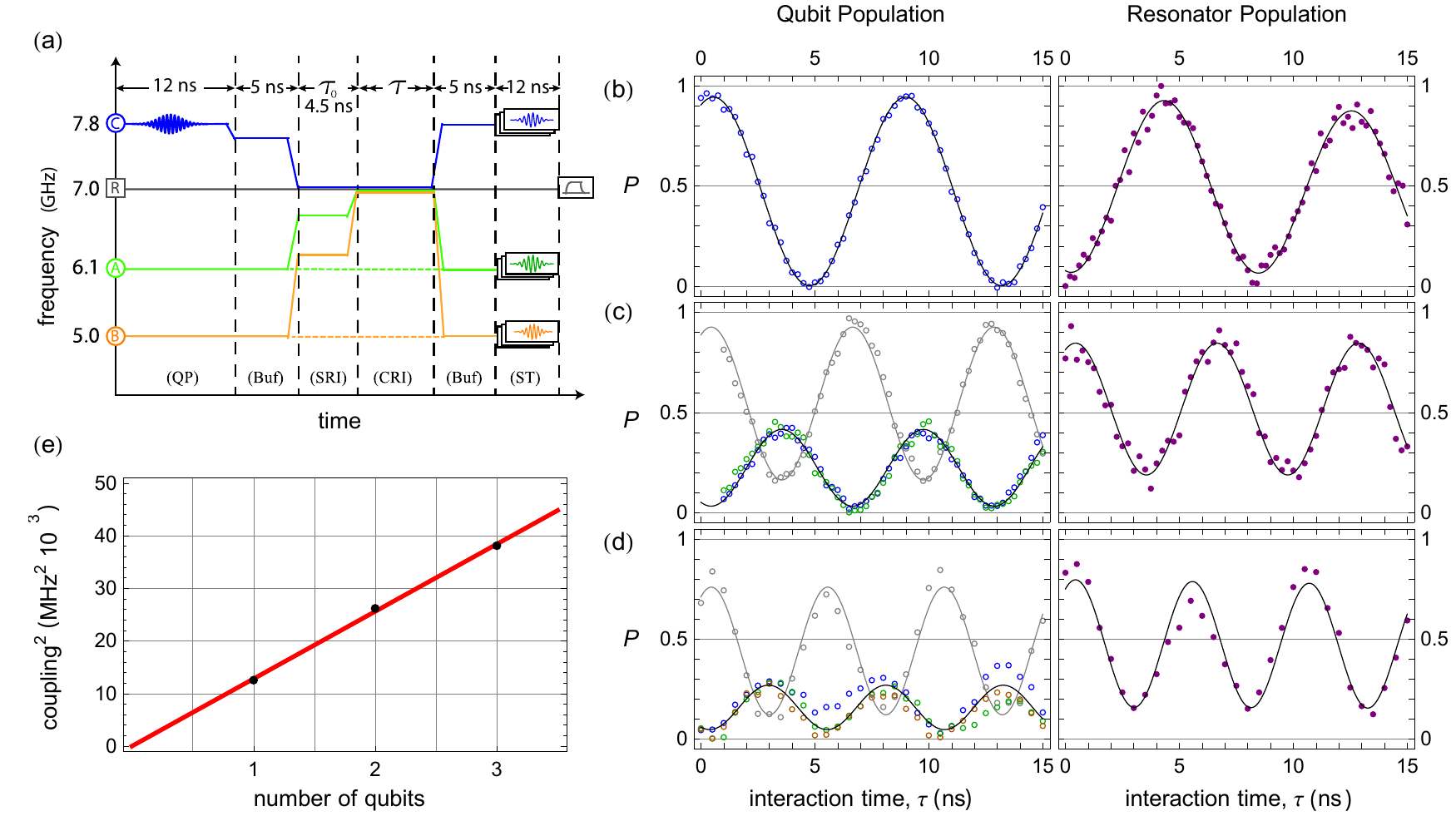}
  \caption{(Color online) Dynamics of the collective vacuum Rabi oscillations. (a) Pulse sequence
  for the single and collective vacuum Rabi oscillation (QP=Qubit Preparation, Buf=Buffer Level for Flux Pulses,
  SRI=Single Resonant Interaction, CRI=Collective Resonant Interaction, ST=State Tomography).
  Qubits not taking part in the collective interaction
  remain at their bias frequency (dashed green/blue). The duration $\tau$ of the collective interaction is varied.
  (b) Oscillations for N=1 (first row) to N=3 (third row). The population is shown for the excited states of qubit A (blue), B (green), C (orange),  the $\ket{ggg}$ state (gray) and the resonator (violet).
  (c) Linear fit to the extracted and squared oscillation frequencies of the resonator population.}
  \label{fig:collosci}
\end{figure*}

The pulse sequence we have implemented to observe the one qubit vacuum Rabi oscillation is depicted in Fig.\,\ref{fig:collosci}(a) considering either one of the qubits. First, a $\pi$-pulse is applied to a single qubit far detuned from the resonator, such that the system can be described by a product state $\ket{e}\otimes\ket{0}$. Then the qubit is tuned into resonance with the resonator using a flux pulse of variable duration $\tau$. To reduce the overshoot the qubit is first tuned to an intermediate buffer frequency. On resonance the energy eigenstates of this system with $n=1$ excitations and $N$ qubits are $\ket{n=1,N=1\pm}=1/\sqrt{2} (\ket{g,1}\pm\ket{e,0})$ and the initial excited qubit state undergoes vacuum Rabi oscillations between $\ket{g,1}$ and $\ket{e,0}$. The frequency of the oscillations is given by $\Omega=2\sqrt{g^2+\Delta^2}$ with $\Omega_0=2g$ for $\Delta=0$. The amplitude of the vacuum Rabi oscillation is maximal on resonance where the excitation is fully exchanged. After the resonant flux pulse of length $\tau$ the qubit is detuned from the resonator again and the energy exchange
process is stopped.
In addition to performing a dispersive readout of the qubit population we determine the resonator population by measuring the average photon number, as given by the time integrated power $\braket{\hat{a}^\dagger \hat{a}}$ at the output of the cavity once the oscillation is stopped \cite{Bozyigit2010}. The qubit population is observed to oscillate with a frequency of $112.0$ MHz, out of phase with the photon field oscillating at $111.2$ MHz [Fig.\,\ref{fig:collosci}(b, left)] and
in good agreement with the spectroscopically obtained value.

We extended the procedure described above as shown in Fig.\ \ref{fig:collosci}(a) to two and three qubits. An initial single photon Fock state is prepared by transferring the full excitation of the first qubit to the resonator by adjusting the interaction time between the qubit and the resonator to $\tau_0=\pi/2g$. Then the second and third qubit are tuned into resonance and the resonant collective interaction proceeds for time $\tau$. All qubit populations, obtained by tomographic state reconstruction for each interaction time $\tau$, are observed to oscillate simultaneously out of phase with the cavity photon number. When the number N of qubits taking part in the resonant interaction is increased, the frequency of the oscillations is observed to scale with $\sqrt{N}$ and the amplitude of the individual qubit population decreases to 1/N fulfilling the normalization condition. The tripartite states between which the oscillations occur are $\ket{g,g,1}$ and $1/\sqrt{2} (\ket{g,e,0}-\ket{e,g,0})$ and equivalently $\ket{g,g,g,1}$ and $1/\sqrt{3}(\ket{g,g,e,0}+\ket{g,e,g,0}-\ket{e,g,g,0})$ for the fourpartite state, where we have denoted the states in a binary order as $\ket{\text{C},\text{B},\text{A},\text{Cavity}}$. The opposite phase of the $\ket{e,g,g,0}$ state originates from the fact that qubit~A has a negative coupling constant.
The qubit populations measured after the collective vacuum oscillations are shown in Fig.\,\ref{fig:collosci}(c/d, left column). The collective oscillation transfers energy back and forth between the qubits and the resonator, such that the population in the cavity is correlated with the population of the qubit $|ggg\rangle$ state. Hence it is instructive to measure the photonic part of the state as done before for the single qubit vacuum Rabi oscillations [Fig. \ref{fig:collosci}(c/d, right column)]. The oscillation frequencies for the resonator population for N=2 and N=3 are $161.8$ MHz =$\,1.05\cdot\sqrt{2}~\overline{g}/2\pi$ and $195.2$\,MHz =$\,1.03\cdot\sqrt{3}~\overline{g}/2\pi$, with $\overline{g}$ being the root mean square \cite{Lopez2007a} of the three spectroscopically obtained coupling constants. These oscillation frequencies clearly demonstrate the $\sqrt{N}$-nonlinearity of the Tavis-Cummings Hamiltonian in time resolved measurements as illustrated in figure \,\ref{fig:collosci}(e). While the collective qubit oscillations are in good agreement with the expected dynamics, we observe a decrease in visibility with increasing N for both the qubit and resonator populations.
We attribute this effect to an imperfect control of the flux pulse amplitudes. Although we have extracted a flux pulse cross coupling matrix at low frequencies, its accuracy is not sufficient to fully compensate the flux cross talk also at high frequencies and to exactly tune all qubits into resonance with the resonator simultaneously.

\begin{figure*}[!t]
\centering
  \includegraphics[width=1.0\textwidth]{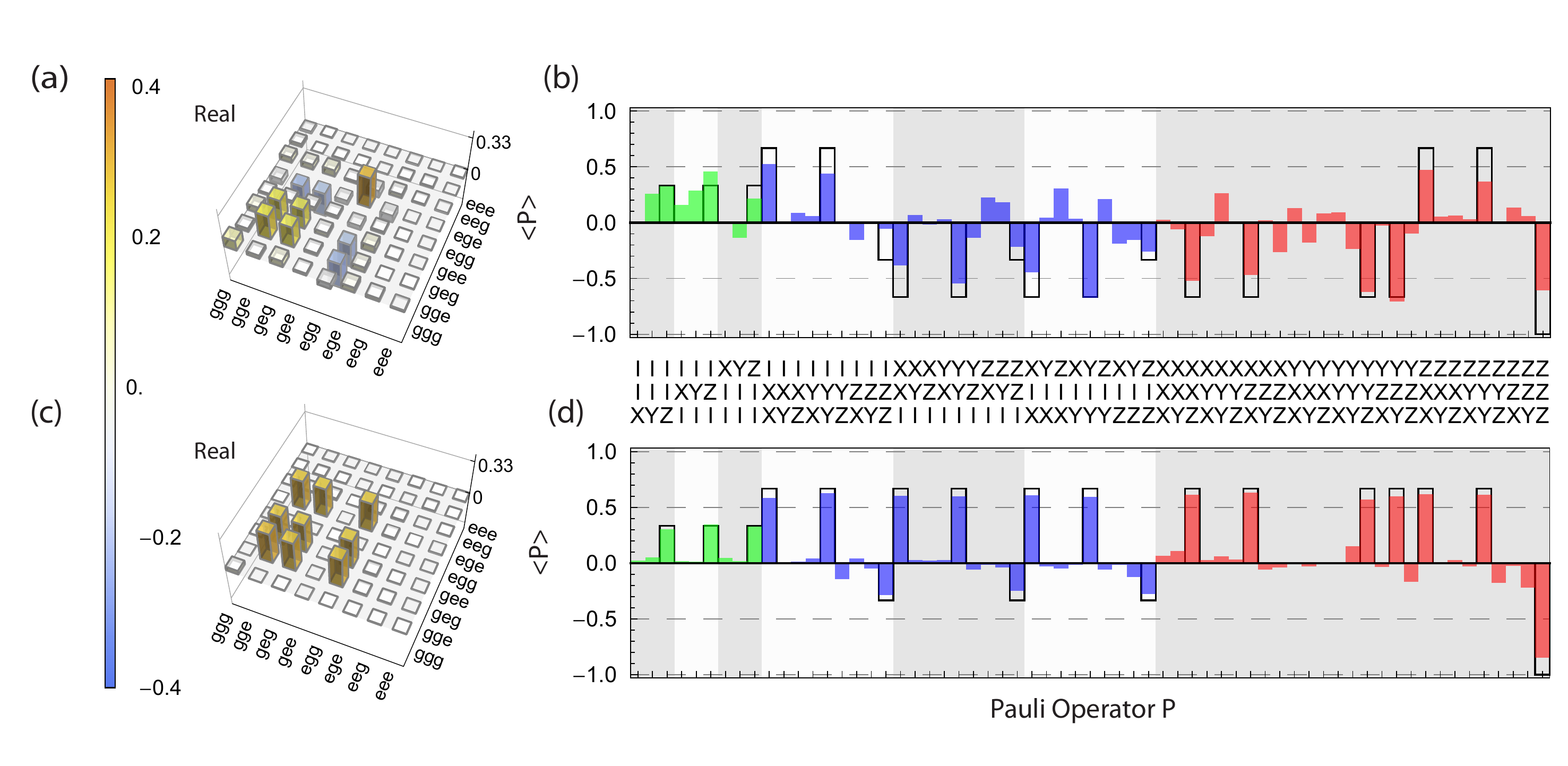}
  \caption{(Color online) Real part of the density matrix of the W-state. (a) Obtained for
  the collective approach as shown in Fig.~\ref{fig:collosci}.
  (c)~Generated by the sequential approach. (b/d) Pauli set for the collective/sequential approach.}
\label{fig:densmat}
\end{figure*}

During the collective oscillations a W-state is created at time $\tau_{W}=\pi/(2\overline{g}\sqrt{N})$, i.e. when the cavity state factorizes $\ket{0} \otimes 1/\sqrt{3} (\ket{g,g,e}+\ket{g,e,g}-\ket{e,g,g})$. We measured the three qubit density matrix at this time by applying full quantum state tomography \cite{Filipp2009b}. Figure\,\ref{fig:densmat} shows (a) the real part of the density matrix and (b) the corresponding Pauli sets.
The collective interaction time needed to create this three qubit entangled state was only $\tau_{W}=\unit[2.9]{ns}$, while the overall sequence duration from the start of the excitation pulse to the end of the flux pulses was \unit[24.4]{ns} as limited by technical constraints.

As needed the dynamic phase that is picked up during the time when the qubits are detuned can be corrected by applying single qubit phase gates or by adjusting the phases of the tomography pulses. Here the coherent entries of the density matrix have been numerically rotated to correspond to the phases in the expected state.

We have also characterized the measured state fidelity defined as $\bra{\rho_t}\rho\ket{\rho_t}$ with respect to the density matrix rotated into the appropriate basis. We use a maximum likelihood method which assumes that the measurement outcomes are subject to Gaussian noise. To find the most probable physical quantum state consistent with the obtained result the problem is mapped to a least square minimization by an appropriate change of the operator basis \cite{Smolin2012}. Using this method the number of steps needed to perform the maximum likelihood algorithm scales as $O(d^4)$, where d is the dimension of the quantum state. The resulting fidelity is 78\,\% with respect to the density matrix $\rho_t=\ket{\Psi_W}\bra{\Psi_W}$ of the ideal state $\Psi_W=1/\sqrt{3} (\ket{g,g,e}+\ket{g,e,g}-\ket{e,g,g})$, limited predominantly by flux cross talk as discussed before in the context of time resolved measurements.
In addition, via convex roof extension we have found the three-tangle to be close to zero $(0.06)$, verifying that the prepared state belongs to the W-class rather then to the GHZ-class \cite{Cao2010}.
Using the entanglement witness operator $\hat{M} = 2/3\,{Id} - \ket{W}\bra{W}$ we find that $Tr(\hat{M} \rho)=2/3 - \mathcal{F} = -0.12 < 0$
allows to discriminate our tripartite entangled pure state against any bipartite entangled states \cite{Guhne2009}.

We note that, in the time resolved measurements the total measurement time per data point is determined  by the $2.5 \, 10^5$ averages per population measurement at a repetition rate of \unit[50]{kHz}. At the same rate each state preparation and subsequent measurement was repeated $6.5 \, 10^5$ times to obtain the density matrix and $10^7$ times to obtain the cavity population.

To experimentally verify that the fidelity of a W-state generated by the collective interaction with a cavity mode in this device is not limited by coherence we have also prepared a W-state using a novel sequential method. In this approach we distribute a single excitation equally between the three qubits using resonant interaction. First, we excite qubit~C and tune it into resonance with the resonator for time $\tau_1=\arcsin{(\sqrt{2/3}\,)}/g_C$. During this interaction, 2/3 of the excitation is transferred to the resonator. Next we tune qubit~B into resonance for a time $\tau_2=\arcsin{(\sqrt{1/2}\,)}/g_B$, transferring half of the resonator excitation to qubit~B. Finally we let qubit~A pick up the remaining third of the energy from the resonator by bringing it into resonance for a time $\tau_3=\arcsin{(1)}/g_A$. The overall sequence duration from the start of the excitation pulse to the end of the flux pulses is \unit[26.7]{ns}, much longer than $\tau_W$ but similar to the total pulse sequence length used for the collective resonant interaction. Using joint readout and tomography we obtain a density matrix of the W-state with much higher fidelity 91\%, shown in figure \ref{fig:densmat}(c/d). As expected the preparation time for the sequential method is longer than for the collective approach, because the latter method exploits the $\sqrt{N}$-enhancement.
The fact that the higher fidelity was obtained by a procedure where no simultaneous flux pulses are applied to multiple qubits affirms that the main limitation of the collective generation is due to residual flux crosstalk, which could be eliminated in future experiments, rather than decoherence. Moreover we estimated the influence of dissipation and dephasing in absence of all systematic errors to obtain a theoretical upper limit of 93\% for the sequential and 97\% for the collective state preparation.

In summary, in a cavity QED experiment, we have studied the collective interaction of up to three superconducting qubits with a single photon stored in a microwave resonator. We have explicitly reconstructed the density matrix of the multi-qubit entangled state of the Dicke- or W-type with fidelities considerably higher than 2/3. We have also resolved the temporal dynamics of the population in the time domain characterized by the $\sqrt{N}$-scaling of the collective vacuum Rabi oscillation frequency in both the qubit and the photon states. The ability to study a collective system while keeping full control over all individual components may allow to study other interesting collective effects such as superradiance or phase multistability in fully controlled small ensembles.

Related work has been carried out independently \cite{lucero2012}.

\begin{acknowledgments}
This work was supported by the Swiss National Science Foundation (SNF) and the EU IP SOLID.
\end{acknowledgments}
%

\end{document}